\documentclass[10pt,conference]{IEEEtran}
\IEEEoverridecommandlockouts

\usepackage{cite}
\usepackage{hyperref}
\usepackage{orcidlink}
\usepackage{comment}
\usepackage{amsmath,amssymb,amsfonts}
\usepackage{algorithmic}
\usepackage{graphicx}
\usepackage{textcomp}
\usepackage{xcolor}
\usepackage{soul}
\usepackage{multicol}
\usepackage{multirow}
\usepackage{subfigure}

\usepackage[nolist]{acronym} 

\def\BibTeX{{\rm B\kern-.05em{\sc i\kern-.025em b}\kern-.08em
    T\kern-.1667em\lower.7ex\hbox{E}\kern-.125emX}}

\begin{document}

\title{Analytical Modeling of Batteryless IoT Sensors Powered by Ambient Energy Harvesting}

\author{\IEEEauthorblockN{Jimmy Fernandez Landivar$^{\mathrm{1}} $\orcidlink{0000-0002-4904-5256}, Andrea Zanella$^{\mathrm{2}} $ \orcidlink{0000-0003-3671-5190}, Ihsane Gryech$^{\mathrm{1}} $ \orcidlink{0000-0001-5288-4205}, Sofie Pollin$^{\mathrm{1}}$\orcidlink{0000-0002-1470-2076}, and Hazem Sallouha$^{\mathrm{1}}$\orcidlink{0000-0002-1288-1023}}
\IEEEauthorblockA{\textit{$^{\mathrm{1}}$Department of Electrical Engineering (ESAT) - WaveCoRE}, KU Leuven, 3000 Leuven, Belgium \\
\textit{$^{\mathrm{2}}$Department of Information Engineering}, University of Padova, 35131 Padova, Italy\\ E-mails: \{jfernand, ihsane.gryech, sofie.pollin, hazem.sallouha\}@esat.kuleuven.be, andrea.zanella@unipd.it}\\

\vspace{-1.2cm}

\thanks{This work was supported by the SwarmNET project under KU Leuven C1 grant No. C14/24/093 and by the SUPERIOT project. The SUPERIOT project has received funding from the Smart Networks and Services Joint Undertaking (SNS JU) under the European Union’s Horizon Europe research and innovation program under Grant Agreement No. 101096021.}
}

\maketitle
\begin{acronym}

\acro{BLE}{Bluetooth Low Energy}
\acro{CL}{Circuit Load}
\acro{EH}{Energy Harvester}
\acro{EHU}{Energy Harvesting Unit}
\acro{ESR}{Equivalent Series Resistance}
\acro{ESS}{Environmental Sensing Service}
\acro{IoT}{Internet of Things}
\acro{IAQ}{Indoor Air Quality}
\acro{MPPT}{Maximum Power Point Tracking}
\acro{MC}{Micro-controller}
\acro{PM}{Power Manager}
\acro{PV}{Photovoltaic}
\acro{RF}{Radio-Frequency}
\acro{RBPi}{Raspberry Pi 4}
\acro{SoC}{System-on-Chip}
\acro{SCap}{Supercapacitor}

\acro{VOC}{Volatile Organic Compound}

\end{acronym}
\begin{abstract}

This paper presents a comprehensive mathematical model to characterize the energy dynamics of batteryless IoT sensor nodes powered entirely by ambient energy harvesting. The model captures both the energy harvesting and consumption phases, explicitly incorporating power management tasks to enable precise estimation of device behavior across diverse environmental conditions. The proposed model is applicable to a wide range of IoT devices and supports intelligent power management units designed to maximize harvested energy under fluctuating environmental conditions. We validated our model against a prototype batteryless IoT node, conducting experiments under three distinct illumination scenarios. Results show a strong correlation between analytical and measured supercapacitor voltage profiles, confirming the proposed model’s accuracy.

\end{abstract}


\section{Introduction}

The large-scale deployment of \ac{IoT} devices is a fundamental requirement to allow accurate and ubiquitous sensing and monitoring in diverse environments~\cite{sallouha2017ulora}. However, as the number of devices deployed increases, so does the number of batteries that contain hazardous substances, limiting the sustainability score of \ac{IoT} networks and introducing environmental concerns \cite{Fernandez24}. To address this, low-power, batteryless IoT sensors that harvest ambient energy reduce battery dependence and environmental impact.

Batteryless \ac{IoT} nodes rely on energy harvested from ambient sources such as \ac{RF}, light, thermal gradients, and kinetic motion \cite{Chatterjee2023ambient}. These nodes leverage \ac{PM} units with highly efficient DC/DC conversion and \ac{MPPT} to make efficient use of the energy harvested from the environment \cite{Ahmad2021mppt}. In large-scale IoT deployments, it is essential to ensure that batteryless \ac{IoT} sensor nodes are carefully designed and optimized before production, which is crucial to maximize energy efficiency and minimize costs \cite{Fernandez24,AZanellaSensors}. This key step in the development of batteryless nodes involves the analysis of ambient energy dynamics in the environment and the detailed understanding of the node's energy behavior. This knowledge enables accurate performance evaluation across diverse environments and operating conditions \cite{Chatterjee2023ambient}. 

Recent works have highlighted the importance of mathematical modeling of energy dynamics in \ac{IoT} devices to optimize energy management and ensure reliable operation under intermittent and varying energy conditions. Focusing on NB-IoT, studies have proposed analytical~\cite{andres2019analytical} and stochastic~\cite{garciamartin2023energy} models that capture device behaviors and energy consumption patterns, addressing critical aspects such as power-saving mechanisms (eDRX, PSM), idle-state transitions, and uplink transmissions. Validated through experiments and NS-3 simulations, these models support battery sizing, lifetime prediction, and parameter tuning, all crucial for energy-efficient \ac{IoT} deployments. Although these efforts focused on battery-powered nodes, they offer valuable insights and underscore the importance of research into energy modeling for \ac{IoT}.  

For batteryless \ac{IoT} nodes employing low-power communication methods like \ac{BLE} and LoRaWAN, analytical modeling has evolved to address the fluctuating nature of energy sources. Sabovic et al.~\cite{savovic2020energyaware} modeled the energy harvesting and management circuit as a current generator, incorporating sensing and communication costs while exploring the effects of sensing frequency and capacitor sizing. Their work, assuming static energy sources and lacking advanced power management, was extended by Delgado et al.~\cite{delgado2021lorawan}. The latter formalized three capacitor-based electrical models, including parasitic effects (ESR, EPR), to predict task completion without power failure. However, their model did not account for real-world dynamic energy use for sensing or complex power management. 
More recently, probabilistic and diffusion-based models~\cite{czachorski2022modelling} have been proposed to better capture non-exponential energy arrivals and consumption patterns, improving predictions of depletion events and system uptime. 
Furthermore, studies on light-powered \ac{IoT} nodes~\cite{onwunali2022modelling} have emphasized the incorporation of empirical irradiation data to optimize harvester sizing and minimize outages.

Together, the state-of-the-art highlights the importance of accurate energy modeling of \ac{IoT} devices within the energy constraints of real-life deployments. However, existing models overlook power management, which is critical to optimizing the operation of batteryless \ac{IoT} nodes and aligning models with practical deployments. This gap drives our work to develop an analytical model that integrates energy management into harvesting and supply processes.

Our contribution expands existing research by introducing a comprehensive mathematical model for batteryless \ac{IoT} sensor nodes powered by dynamic ambient energy. We begin with an in-depth analysis of the \ac{EHU} (comprising the ambient \ac{EH}, \ac{PM} module, and a \ac{SCap}-based energy buffer) and subsequently model the energy dynamics of the \ac{CL}. These components are integrated into a unified model that captures the interactions between active and sleep cycles, enabling simulation of energy flow and voltage variations in the energy buffer. This model is generalizable to a wide range of batteryless systems, enabling energy profiling and performance optimization before physical deployment.

The main contributions of this work are twofold:
\begin{itemize}
    \item We analyze a batteryless \ac{IoT} sensor node designed for environmental sensing, describing its architectural components and developing a mathematical model of its dynamic energy variations. This model uses an electrical circuit representation of the \ac{EH}, \ac{PM}, \ac{SCap}, \ac{CL}, along with the node's functional states (sensing, communication, and sleeping). We also detail the energy charge/discharge behavior of the \ac{SCap} in each of these states.
    \item We validate the model using experimental data from a light-powered batteryless \ac{IoT} prototype, assessing the correlation between simulated and real-world results under three lighting conditions (300~lx, 500~lx, and 700~lx), including variations to show when the energy consumed by the node is either equal to, greater than, or less than the energy harvested. Our results demonstrate that the model accurately predicts the real-world \ac{SCap} energy behavior, with deviations of around 1\% in energy.
\end{itemize}


\section{Components of a Batteryless IoT Sensor}
\label{Sec_II}
\begin{figure}[t]
\centering
\includegraphics[scale=0.28]{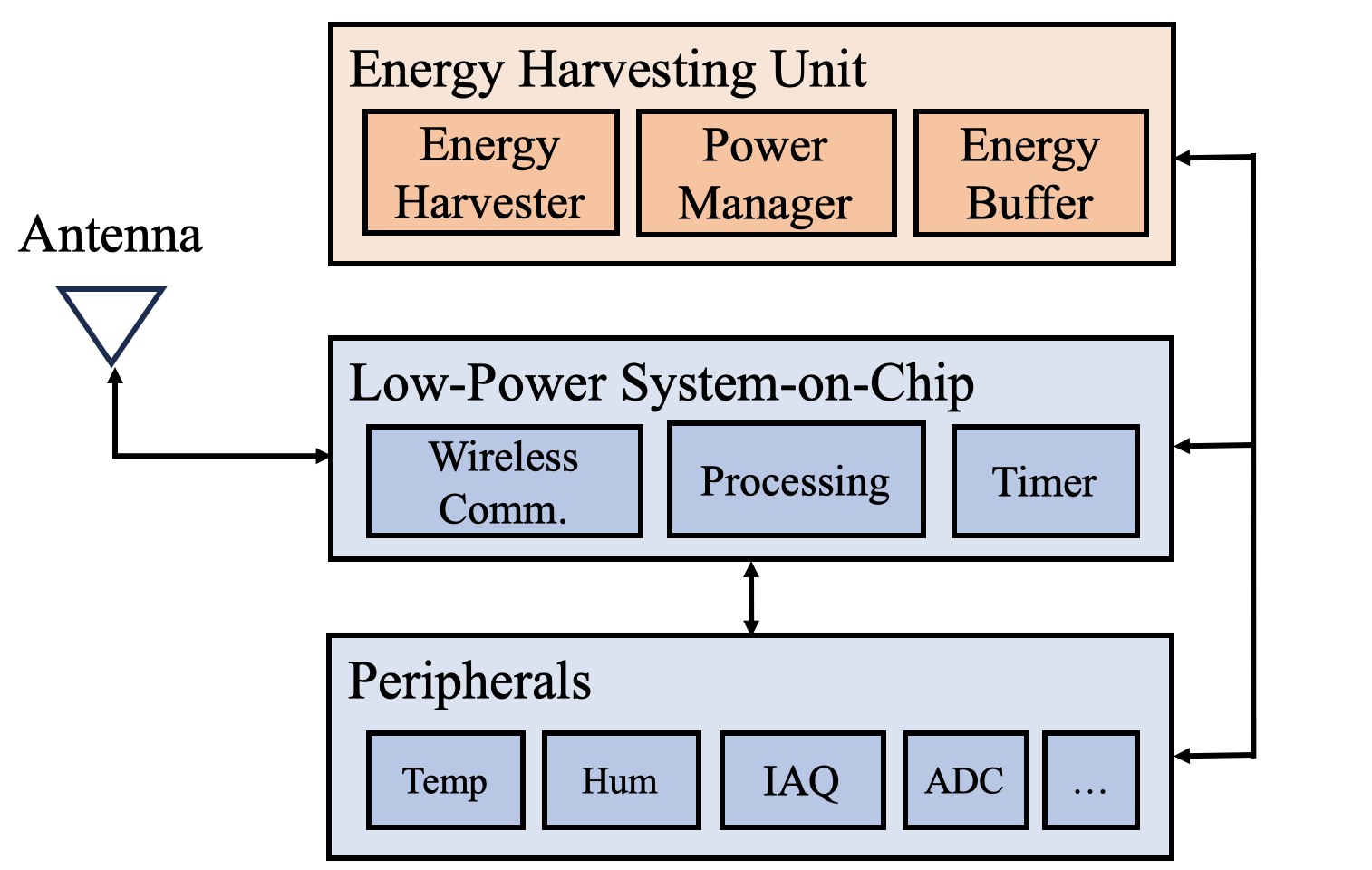}
\vspace{-0.5em}
\caption{Principal components of a batteryless \ac{IoT} sensor node.}
\vspace{-1.2em}
\label{fig:net_diagram}
\end{figure}

The block diagram in Fig. \ref{fig:net_diagram} depicts the principal components of a batteryless \ac{IoT} device designed for sensing and monitoring. These components include the \ac{EHU}, a low-power \ac{SoC} with integrated radio circuits and the different low-power sensors. In the following, we provide a detailed description of the main characteristics of each component and the energy constraints necessary to achieve continuous batteryless operation.

\subsection{Batteryless IoT Sensor Node Components}

\subsubsection{Energy Harvesting, Storage and Supply}
The \ac{EHU} harvests ambient energy (e.g., indoor \ac{RF}, light, thermal) and manages its storage in an energy buffer, typically a \ac{SCap}. It includes two main parts: the \ac{EH}, such as a \ac{PV} cell for light harvesting, and the \ac{PM}, which efficiently transfers energy to the \ac{SCap} and supplies the \ac{CL}. To ensure stable operation despite variable energy sources, the \ac{PM} uses \ac{MPPT} techniques to maximize harvested power. Regarding the energy buffer, \ac{SCap}s are favored over conventional batteries due to their high charge-discharge efficiency (97\%-98\%), nearly unlimited cycles, and minimal thermal dissipation, making them ideal for energy-constrained~\ac{IoT}~\cite{2529}.

\subsubsection{Low-power System-on-Chip}
A \ac{SoC} is an integrated circuit that consolidates various system components such as memory, analog and digital peripherals, and connectivity interfaces into a single substrate, managed by a microprocessor. Communication within the \ac{SoC} can occur via wired protocols like I\textsuperscript{2}C, SPI, or UART, as well as through wireless technologies such as Wi-Fi and \ac{BLE}. \ac{SoC}s provide a compact and efficient solution for sensing, processing, and communication tasks in embedded systems. Among the most widely adopted options are \ac{SoC}s based on the ARM Cortex-M series, specifically designed for low-power applications \cite{Botirov2023,Fernandez24}.

\subsubsection{Low-Power Sensors and Peripherals}
To achieve accurate and efficient sensing while minimizing energy consumption, a batteryless \ac{IoT} node integrates low-power sensors specifically designed for each application. The state-of-the-art sensors are capable of measuring key environmental parameters with one integrated multipurpose all-in-one sensor. For example, in \ac{IAQ}, the sensor includes temperature, humidity, air pressure, and \acp{VOC}. By utilizing an accurate and energy-efficient all-in-one sensing approach, the system ensures prolonged operation using only the harvested energy \cite{Fernandez24}.

\subsection{Energy Constraints for Continuous Batteryless Operation}
\label{II.1}
In our previous work \cite{Fernandez24}, we defined the critical energy constraints necessary for designing reliable batteryless \ac{IoT} sensor nodes capable of continuous operation. We considered a constant illumination scenario to guarantee continuous operation and sensor readings at a fixed amount of time. Moreover, to achieve maximum sensing efficiency with the available energy, our solution adopted a power management strategy based on the periodic switching between active and sleep cycles. Our previous analysis considers all the energy sources and energy consumed by the batteryless \ac{IoT} node and, based on the energy conservation principle, outlines that the energy produced in a total operational time $T$ should be greater than or equal to the energy consumed, which is described in the following inequality: 
\begin{equation}
E_{buf}\!+\!\underbrace{\int_{0}^{T}\!P_{harv}(\tau)d\tau}_{E_{harv}(T)}\geq\underbrace{\int_{0}^{T}\!P_{dev}(\tau)d\tau}_{E_{dev}(T)},
\label{eq1}
\end{equation}
where $\tau$ is used as the time variable and $E_{buf}$ represents the minimal energy in the buffer (in Joules), to ensure that the node stays operational. $P_{harv}(\tau)$ is the harvested power for an instant of time $\tau$ in Watts, and $P_{dev}(\tau)$ is the power consumed by the node for an instant of time $\tau$ in Watts. In addition, $E_{harv}(T)$ and $E_{dev}(T)$ represent the energy harvested and consumed (in Joules), respectively, during a complete duty cycle denoted by $T$. A simplified version of the energy flow model~is
\begin{equation}
E_{harv}(T)\geq\!E_{dev}(T), \forall T \in [0, \infty ).
\label{eq2}
\end{equation}
In other words, at each duty cycle, we aim to harvest more or equal energy than what the node consumes for the active and sleep cycles. Therefore, the inequality described in \eqref{eq2} models the energy of an \ac{IoT} node that is always active. However, to provide a detailed view of the energy produced and consumed during sleep and active cycles, we divide the duty cycle time $T$ into active and sleeping times, $T_a$ and $T_s$, such that $T$ = $T_a$ + $T_s$. Consequently, we divide the energy consumed by the node $E_{dev}(T)$ into energy consumed during active and sleeping times, $E_{deva}(T_a)$ and $E_{devs}(T_s)$, being $E_{dev}(T)= E_{deva}(T_a) + E_{devs}(T_s)$. Therefore, the energy conservation of a batteryless \ac{IoT} node with low-power management is modeled as
\begin{equation}
E_{harv}(T_a+T_s)\geq\!E_{deva}(T_a)+E_{devs}(T_s).
\label{eq5}
\end{equation}
The obtained mathematical expression is applied to design and implement batteryless \ac{IoT} sensor nodes under constant illumination scenarios. Specifically, it allows us to compute the sleeping period $T_s$ that satisfies \eqref{eq5}, enabling an active/sleep low-power management strategy for the batteryless \ac{IoT} nodes. This will allow us to evaluate and model the energy dynamics of the node in different illumination scenarios.

\section{Energy Modeling of Batteryless IoT Sensors}
\label{Sec_III} 

To accurately model the behavior of batteryless \ac{IoT} sensor nodes powered by ambient energy harvesting, we first analyze their energy harvesting and consumption stages. This enables a detailed characterization of node performance and supports the development of a mathematical model capturing its real-world operation. While this work focuses on \ac{BLE}-enabled environmental sensing nodes, the model is generalizable to any batteryless \ac{IoT} device, offering valuable insights for network optimization and design.

\begin{figure}[th]
\centering
\includegraphics[scale=0.2]{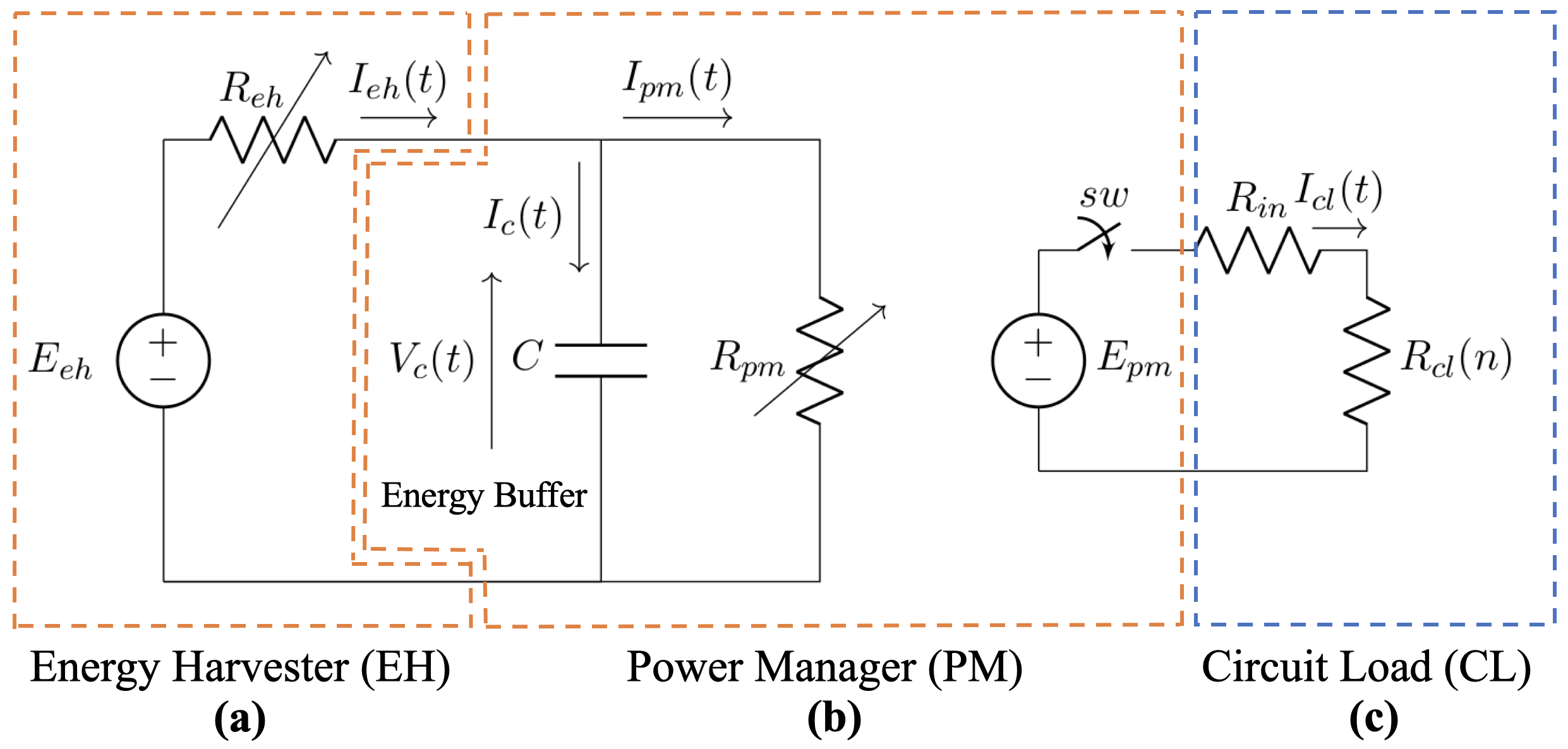}
\vspace{-1em}
\caption{Equivalent electrical model of the node: the left-hand side circuit models the behavior of the voltage at the \ac{SCap}, while the right-hand side circuit models the \ac{CL}.}
\vspace{-1em}
\label{fig:model}
\end{figure}

The analysis begins with the \ac{EH}, which converts ambient energy (\ac{RF}, light, thermal, kinetic) into electrical energy following a technology-specific transduction curve~\cite{Chatterjee2023ambient,epishine2024evalkit}. The \ac{PM} then manages this power via DC/DC conversion and \ac{MPPT}, optimizing energy transfer under dynamic conditions~\cite{Ahmad2021mppt}. Harvested energy is stored in a \ac{SCap}, which buffers and stabilizes supply to the \ac{CL}. The \ac{PM} regulates the voltage and current delivered to the \ac{CL} to ensure stable operation. Finally, the \ac{CL} consumes energy based on the node's actions determined by the different \textit{states} of the finite state machine governing the node's behavior. In this section, we first model each component individually, then derive the governing equations for currents and voltages across the circuit model, depicted in Fig.\ref{fig:model}.

\subsection{Description of the circuit components}

\paragraph{\ac{EH} and PM}
The \ac{EH} gathers energy from the environment to charge the \ac{SCap} and power the \ac{CL}. The harvester can be modeled as an ideal constant current source with a parallel inner resistance. The current generated by the \ac{EH}, as well as the output voltage, depend on the specific technology used to harvest energy from the environment and on the environmental conditions. For example in light, the energy harvested from a \ac{PV} cell depends on the light (50 - 1000 lx), temperature, and humidity. The \ac{PM} adapts the load connected to the \ac{EH} to guarantee maximum power transfer to the circuit, i.e., to the \ac{SCap} and \ac{CL}. Moreover, the \ac{PM} controls the maximum voltage of the \ac{SCap}.

The combination of \ac{EH} and \ac{PM} can be modeled as the series of an ideal voltage generator, with nominal voltage $E_{eh}$, and a resistance $R_{eh}$, which varies with the voltage $V_c(t)$ at the \ac{SCap}. Accordingly, the overall power transferred to the circuit remains equal to $P_{eh}(t)$. Mathematically, we have
\begin{equation}
P_{eh}(t) = V_c(t) I_{eh}(t) =  V_c(t)\frac{E_{eh}-V_c(t)}{R_{eh}}\,,
\end{equation}
and hence $R_{eh}$ can be expressed as
\begin{equation}
R_{eh} = V_c(t)\frac{E_{eh}-V_c(t)}{P_{eh}(t)} \,,
\label{eq:Reh}
\end{equation}
where $I_{eh}(t)$ is the current generated by $E_{eh}$ and $V_c(t)$.
It is important to note that when $P_{eh}(t)=0$ or $V_c(t)=E_{eh}$, the model is no longer valid because, in these cases, the \ac{PM} will disconnect the \ac{EH} from the \ac{SCap} to prevent inefficient power transfer or potential overcharging. We model these two limit situations by setting $E_{eh}=0$ and $R_{eh}=R_c$, where $R_c$ is the leakage 
resistance of the \ac{SCap} that accounts for the self-discharging of the \ac{SCap} in the absence of harvested power. 

\paragraph{SCap} The \ac{SCap} is modeled as an ideal capacitor with capacitance $C$ that closes the circuit. An additional parallel resistance $R_c$ models the self-discharging in the absence of harvested power. The current $I_c(t)$ is positive when the \ac{SCap} is charging, i.e., storing energy from the \ac{EH}, and negative when the \ac{SCap} is discharging, i.e., powering the \ac{CL}.  

\paragraph{PM and \ac{CL}}
The resistance \ac{CL} on the right-hand side circuit represents the equivalent load of the other components of the boards, including \ac{MC}, radio transceiver, and sensors, and its value depends on the operations performed by these components. This resistance dissipates a power $P_{cl}$ that depends on the operational state of the node. Since a node can only operate in a finite number of states, the power $P_{cl}$ typically takes values in a finite set $\mathcal{P}_{cl}=\{\ell_0,\dots,\ell_{M-1}\}$.

For proper operation, digital components typically require a stable voltage source. The role of the \ac{PM} is to stabilize the voltage used to power the \ac{CL}, fixing it to a nominal value $E_{pm}$. We model it as a quadripole, where the left-hand side circuit consists of a variable resistance $R_{pm}$ and the right-hand side circuit consists of an ideal voltage generator $E_{pm}$. This is modeled in series with its inner resistance $R_{in}$, and a switch $sw$, which controls the energy supply to the \ac{CL}. 

The resistance $R_{pm}$ on the right-hand-side circuit represents the equivalent load presented by the \ac{PM} to the \ac{EH} and the \ac{SCap}, whereas the inner resistance $R_{in}$ on the right-hand side circuit accounts for the power loss due to voltage scaling. Therefore, the power $P_{pm}$ dissipated by $R_{pm}$ is expressed as
\begin{eqnarray}
P_{pm} &=& P_{cl} w_{pm}, \quad\hbox{with} \quad w_{pm} = \frac{1}{1+\frac{R_{in}}{R_{cl}}}\,.
\label{eq:Ppm}
\end{eqnarray}
where the coefficient $w_{pm}\geq 1$ accounts for the inefficiency of voltage scaling, which is a characteristic of the \ac{PM}. 

In a certain state of the \ac{CL}, such as sleeping, $ P_{cl}$ is constant because of the fixed energy consumption $E_{devs}$. On the other hand, the power dissipated by the resistance $R_{pm}$ is given by
\begin{equation}
P_{pm} = \frac{V_c(t)^2}{R_{pm}}\,.
\end{equation}
Therefore, to have $P_{pm}=P_{cl}w_{pm}$ the resistance $R_{pm}$ needs to vary with the voltage $V_c(t)$ as follows
\begin{equation}
R_{pm} = \frac{V_c(t)^2}{P_{pm}}=\frac{V_c(t)^2}{w_{pm}P_{cl}}\,.
\label{eq:Rpm}
\end{equation}
Eq.~\eqref{eq:Rpm} provides the value of the equivalent resistance $R_{pm}$ for a given operational state of the circuit, when varying the voltage of the capacitor. 

It is worth noting that if the voltage at the \ac{SCap} (which is proportional to the stored energy) drops below a certain threshold $V_{\rm off}$, the \ac{PM} will no longer be able to keep the output voltage to $E_{pm}$, and it will switch to an OFF state. In this state, the \ac{CL} is no longer powered, and the electronic components of the board (\ac{MC}, transceiver, sensors, volatile memories, etc) will be shut off. In the OFF state, the switch in the right-hand side circuit is open ($sw=0$), which means $P_{pm}=0$,  and consequently $R_{pm}=\infty$.

In general, the reactivation of the circuit after an OFF state may involve a transient period before re-establishing the normal operational conditions. In this period, the node may consume a non-negligible amount of energy, according to the characteristics of the board \cite{Fernandez24,epishine2024evalkit}. 

\subsection{SCap voltage behavior over time }
In this subsection, we derive the mathematical expressions that govern the electrical circuit in the different states (sensing, communication, and sleeping), focusing in particular on $V_c(t)$, i.e., the behavior over time of the \ac{SCap} voltage. Our analysis assumes $P_{eh}(t)=p_h$ and $R_{cl}(t)=\rho_h$ for $t \in [0,T)$, for an operational time $T$. In other words, we assume the charging and load conditions of the circuit remain fixed in an operational time interval. Moreover, we denote by $V_c(0)$ the voltage of the capacitor at the beginning of the operational time interval. 

In the following, we consider all possible configurations that require a specific analysis. 

\subsubsection{OFF state, EH disconnected}
In OFF state, the right-hand side circuit does not consume energy. Consequently, $I_{pm}$ is zero, which implies $\rho_h = \infty$. 
If the \ac{EH} does not produce enough energy, the \ac{PM} disconnects it from the \ac{SCap}, so that $p_h=0$, while $R_{eh}=R_c$. In this condition, the \ac{SCap} slowly discharges over time, due to the leakage current. The voltage hence varies in time according to the standard discharging law of a capacitor with time constant $\tau_{disch}~=~C\cdot R_c$: 
\begin{equation}
V_c(t) = V_c(0) e^{-\frac{t}{CR_c}}\,.
\end{equation}

\subsubsection{OFF state, EH connected}
If the \ac{PM} is in OFF state but the energy provided by the \ac{EH} is positive, $p_h>0$, then the \ac{SCap} gets charged according to the following differential equation:
\begin{equation}
\frac{{\rm d}V_c(t)}{{\rm d}t} = \frac{I_c(t)}{C} =  \frac{1}{C}\frac{E_{pm}-V_c(t)}{R_{eh}}
=  \frac{1}{C}\frac{p_h}{V_c(t)} \;, %
\end{equation}
where, using \eqref{eq:Reh} and solving the differential equation, we get 
\begin{equation}
V_c(t) = \sqrt{\frac{2 p_h t}{C} + V_c(0)^2} \,.
\end{equation}
It is important to note that if the circuit configuration remains unchanged, the capacitor voltage will continue to rise until it reaches the threshold $V_{on}$. At this point, the \ac{PM} switches are in the ON state, powering the \ac{CL}, and the behavior of $V_c(t)$ will change accordingly. 

\subsubsection{ON state, EH disconnected}
We now consider the situation where the \ac{PM} is in ON state, so that the \ac{CL} dissipated a certain power $P_{cl}$.  
If $p_{h}=0$, the energy required to power the \ac{CL} is taken from that stored in the \ac{SCap}, which is connected in parallel to $R_c$ and $R_{pm}$. The voltage then changes according to the following differential equation

\begin{align}
\frac{{\rm d}V_c(t)}{{\rm d}t} &= \frac{I_c(t)}{C} = -\frac{V_c(t)}{C} \left( \frac{1}{R_c} + \frac{1}{R_{pm}} \right) \notag \\
&= -\frac{V_c(t)}{C} \left( \frac{1}{R_c} + \frac{w_{pm} P_{cl}}{V_c(t)^2} \right).
\end{align}

where, once again, we replace $R_{pm}$ with its expression given by \eqref{eq:Reh}. Solving the differential equation we find 
\begin{equation}
 V_c(t)  = \sqrt{\alpha e^{-2t/(CR_c)}-w_{pm}P_{cl}R_c} 
\label{Vcdisch}
\end{equation}
where $\alpha = V_c(0)^2+w_{pm}P_{cl}R_c\,$. To be observed that \eqref{Vcdisch} holds for $t\leq \frac{CR_c}{2}\ln\left(\frac{\alpha}{w_{pm}P_{cl}R_c}\right)$, since the voltage of the \ac{SCap} cannot become negative. 

\subsubsection{ON state, EH connected}
This is the normal operational condition of the node, where the load dissipates power $P_{cl}$ and the \ac{EH} transfers a positive power $p_h>0$ to the circuit. In this case, the voltage generator on the left-hand side circuit provides a constant voltage $E_{eh}$, while the series resistance $R_{eh}$ varies as for \eqref{eq:Reh}. 
The differential equation that determines the time evolution of the \ac{SCap} voltage is then given by
\begin{equation}
\frac{{\rm d}V_c(t)}{{\rm d}t} = \frac{1}{C}\left(\frac{E_{eh}-Vc(t)}{R_{eh}}-\frac{V_c(t)}{R_{pm}}\right)
=\frac{P_{eh}-w_{pm}P_{cl}}{CV_{c}(t)} \,,
\label{eq:Pon}
\end{equation}
where the right-most expression is obtained by replacing $R_{eh}$ and $R_{pm}$ with their expressions given by \eqref{eq:Reh} and \eqref{eq:Rpm}, respectively. From \eqref{eq:Pon}, we can immediately realize that, when the power provided by the \ac{EH} exceeds that required by the \ac{CL}, the \ac{SCap} will accumulate the extra energy, while in the contrary case, the energy stored in the \ac{SCap} will be used to meet the demand of the \ac{CL}. Solving \eqref{eq:Pon} we get
\begin{equation}
V_c(t) = \sqrt{2\frac{P_{eh}-w_{pm}P_{cl}}{C}t } + V_c(0)^2\,.
\end{equation}
Here, $V_c(t)$ is the voltage in the \ac{SCap} at any time $t$ during the operational time $T$, when the \ac{EH} and/or the \ac{SCap} are supplying energy to the \ac{CL}. This expression allows us to model in detail the energy in the \ac{SCap}, precisely its charge because of energy harvesting and its discharge due to the \ac{CL}.

\section{Experimental Performance Evaluation}
\label{Sec_IV}

To evaluate the proposed mathematical model, we compare its performance with energy measurements from our batteryless sensor node prototype for environmental monitoring with \ac{BLE} communication, introduced in our work in~\cite{Fernandez24}. The node harvests energy from indoor light and periodically senses and transmits data, including \acp{VOC} for \ac{IAQ}, air pressure, temperature, and humidity. Communication with the gateway is bidirectional, enabling both data transmission and command reception. While the gateway in~\cite{Fernandez24} was based on a standalone \ac{SoC}, in this work, it has been upgraded to a Raspberry Pi 4 single-board computer, which collects sensor data and stores it in a structured SQL database, facilitating easier data access and supporting network scalability.

The node hardware integrates an Epishine LEH3 indoor light \ac{EHU}, a Bosch BME680 all-in-one environmental sensor, and a Seeed XIAO \ac{BLE} development board based on the ARM Nordic nRF52840 \ac{SoC} with Bluetooth~5.0 support. The \ac{EHU} includes an e-Peas AEM10941 \ac{PM} with \ac{MPPT} and a 400\,mF CAP-XX GA230F \ac{SCap} for energy storage. On the software side, energy-efficient algorithms, built on top of the manufacturer's firmware with open-source libraries, manage sensor acquisition, structure the data into arrays, and transmit it using the \ac{BLE} Environmental Sensing Service (ESS). Data transmission begins when the node advertises the ESS service for up to 4 seconds at a transmission power of 4\,dBm. Upon connection, the gateway requests the ESS attributes containing the sensor data. Finally, the node either receives gateway commands or enters a sleep cycle.

\begin{figure}[t]
    \centerline{\includegraphics[scale=0.193]{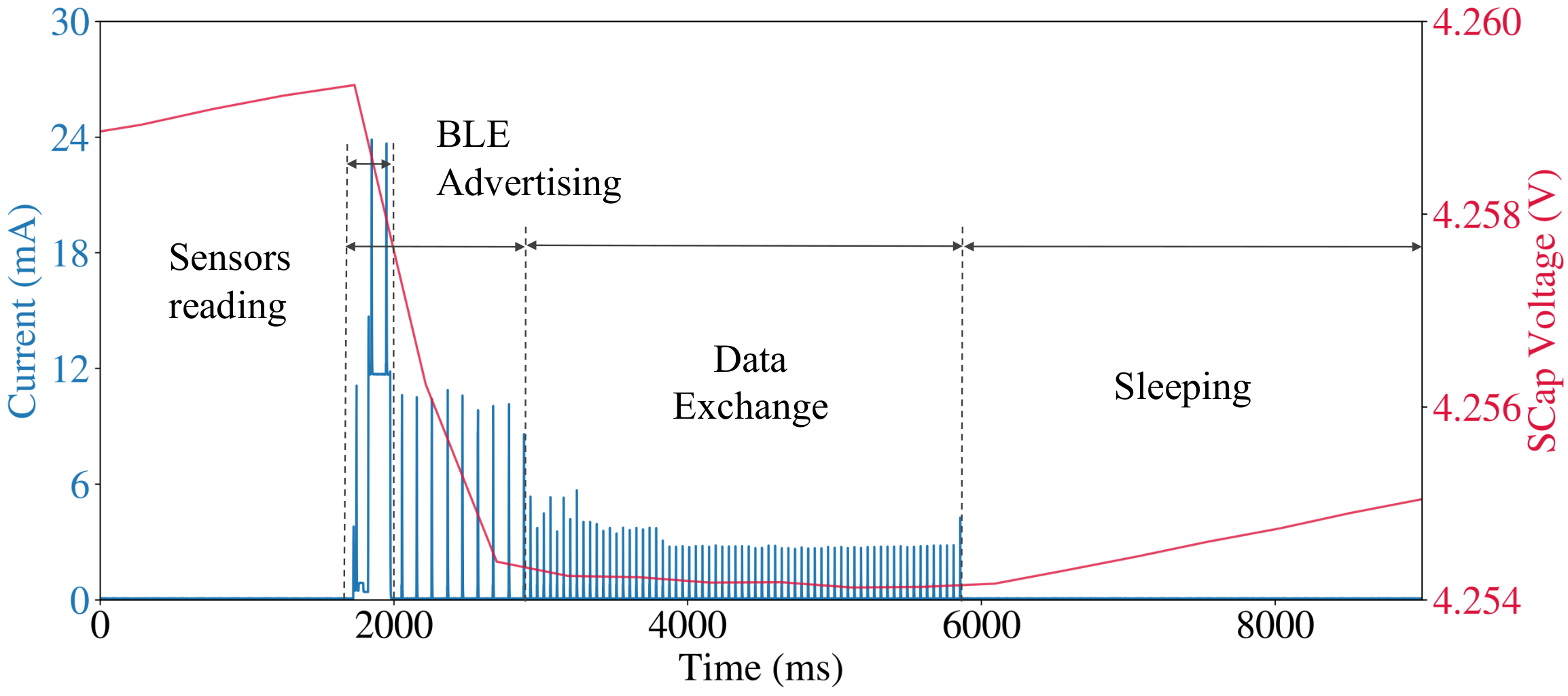}}
    \vspace{-1em}
    \caption{Batteryless IoT node energy consumption profile.}
    \vspace{-1.5em}
     \label{fig:all_sensorsBLE}
\end{figure}

In our experiments, we measure the node's current drawn in each operational state (sensing, \ac{BLE} communication, and sleeping). Subsequently, considering an operational voltage of $3.3~V$, and given the produced current of the \ac{EHU} \cite{epishine2024evalkit}, we computed the active and sleep cycles for three illumination levels by characterizing the node's energy consumption and applying \eqref{eq5}, as described in subsection~\ref{II.1}. The corresponding current levels, summarized in Table~\ref{tab:BLEenergy} and illustrated in Fig.~\ref{fig:all_sensorsBLE} together with the \ac{SCap} voltage, define the energy consumption profile of the node. Table~\ref{tab:BLEenergy} includes the computation of the sleeping time $T_s$ for 300 lx, 500 lx and 700 lx.

\begin{table}[t]
\vspace{-0.2em}
\caption{Batteryless \ac{IoT} Node Energy Consumption Profile}
\vspace{-0.5em}
\centering
\fontsize{7}{10}\selectfont 
\begin{tabular}{|cl|c|c|c|}
\hline 
\multicolumn{1}{|l|}{}                                                                                  & \textbf{Operational State} & \textbf{Current (mA)} & \textbf{Time (s)} & \textbf{Energy (J)} \\ \hline \hline
\multicolumn{1}{|c|}{\multirow{3}{*}{\textbf{\begin{tabular}[c]{@{}c@{}c@{}}Active \\ Cycle\\$E_{deva}$\end{tabular}}}} & Sensors Reading          & 7.550                     & 0.260                 & 0.0065                   \\ \cline{2-5} 
\multicolumn{1}{|c|}{}                                                                                  & BLE Advertising          & 0.400                     & 2.000                 & 0.0026                   \\ \cline{2-5} 
\multicolumn{1}{|c|}{}                                                                                  & Data Exchange            & 0.225                     & 3.000                 & 0.0022                   \\ \hline
\multicolumn{2}{|c|}{\textbf{Sleep Cycle $E_{devs}$ 300lx}}                                                                                         & 0.070                     & 209.9                 & 0.0484                   \\ \hline
\multicolumn{2}{|c|}{\textbf{Sleep Cycle $E_{devs}$ 500lx}}                                                                                         & 0.070                     & 42.44                 & 0.0098                   \\ \hline
\multicolumn{2}{|c|}{\textbf{Sleep Cycle $E_{devs}$ 700lx}}                                                                                         & 0.070                     & 18.10               & 0.0041                   \\ \hline
\end{tabular}
\label{tab:BLEenergy}
\vspace{-0.5em}
\end{table}

The experiments performed to validate our proposed model, comprise a simulation campaign alongside prototype measurements under three constant illumination levels: 300~lx, 500~lx, and 700~lx, with a measurements error of $\pm 30$~lx.
We implement the proposed energy model in MATLAB, configuring all parameters based on the detailed empirical characterization of the prototype node's current consumption. The simulation is configured with illumination scenarios matching the experiments, using energy harvesting data from the manufacturer~\cite{epishine2024evalkit} and lab-based characterization. Additionally, for each of the three considered illumination levels, we perform three tests by adjusting the light intensity by $\pm 100$~lx, creating conditions where $(a)~E_{dev} = E_{harv}$, $(b)~E_{dev} < E_{harv}$, and $(c)~E_{dev} > E_{harv}$. During both simulation and experimental campaigns, the \ac{SCap} voltage $V_c(t)$ was continuously monitored. Results are compared to assess the accuracy of the proposed model by analyzing the \ac{SCap} charge/discharge behavior under varying energy harvesting conditions.

\begin{figure}[t]
    \vspace{-0.1em}
    \centerline{\includegraphics[scale=.235]{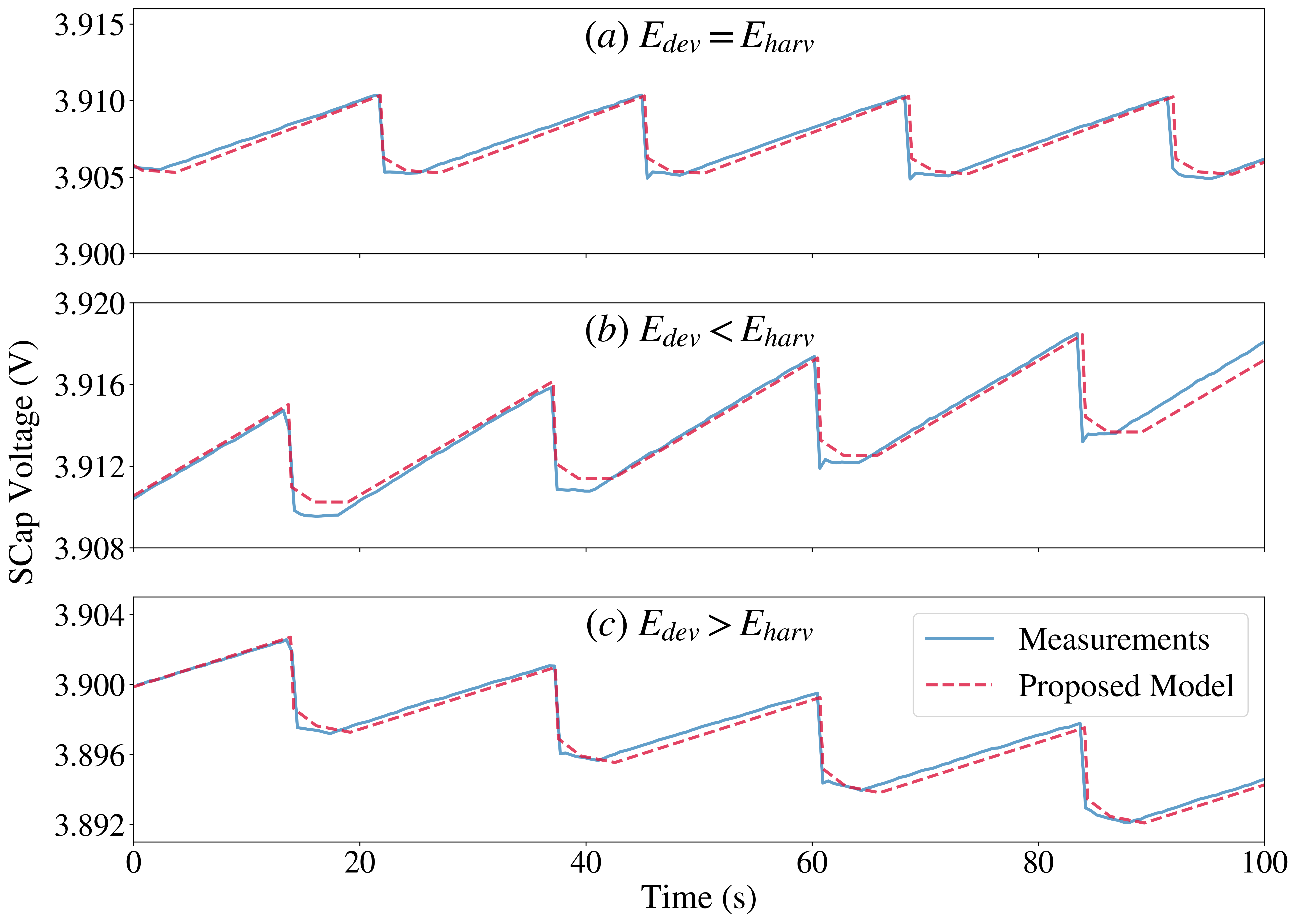}}
    \vspace{-1em}
    \caption{Performance results under the duty cycle of 700~lx.}
    \label{fig:SCapV700lx}
    \vspace{-0.5em}
\end{figure}

Figure~\ref{fig:SCapV700lx} shows the results under 700~lx illumination, where the node operates with a duty cycle ($T_{a}+T_{s}$) of 23.3~s. The high harvested energy rendered small illumination fluctuations negligible, enabling stable operation and a strong match between simulation and measurements across all tests. For example, in Fig.~\ref{fig:SCapV700lx}~(a), minor discrepancies appear in the active cycle due to the BLE advertising interval, modeled as fixed (2~s) but random (0–4~s) in practice. However, as shown in the figure, this discrepancy does not affect the proposed model accuracy, as charge/discharge cycles align closely.

\begin{figure}[t]
    \vspace{-0.5em}
    \centerline{\includegraphics[scale=.235]{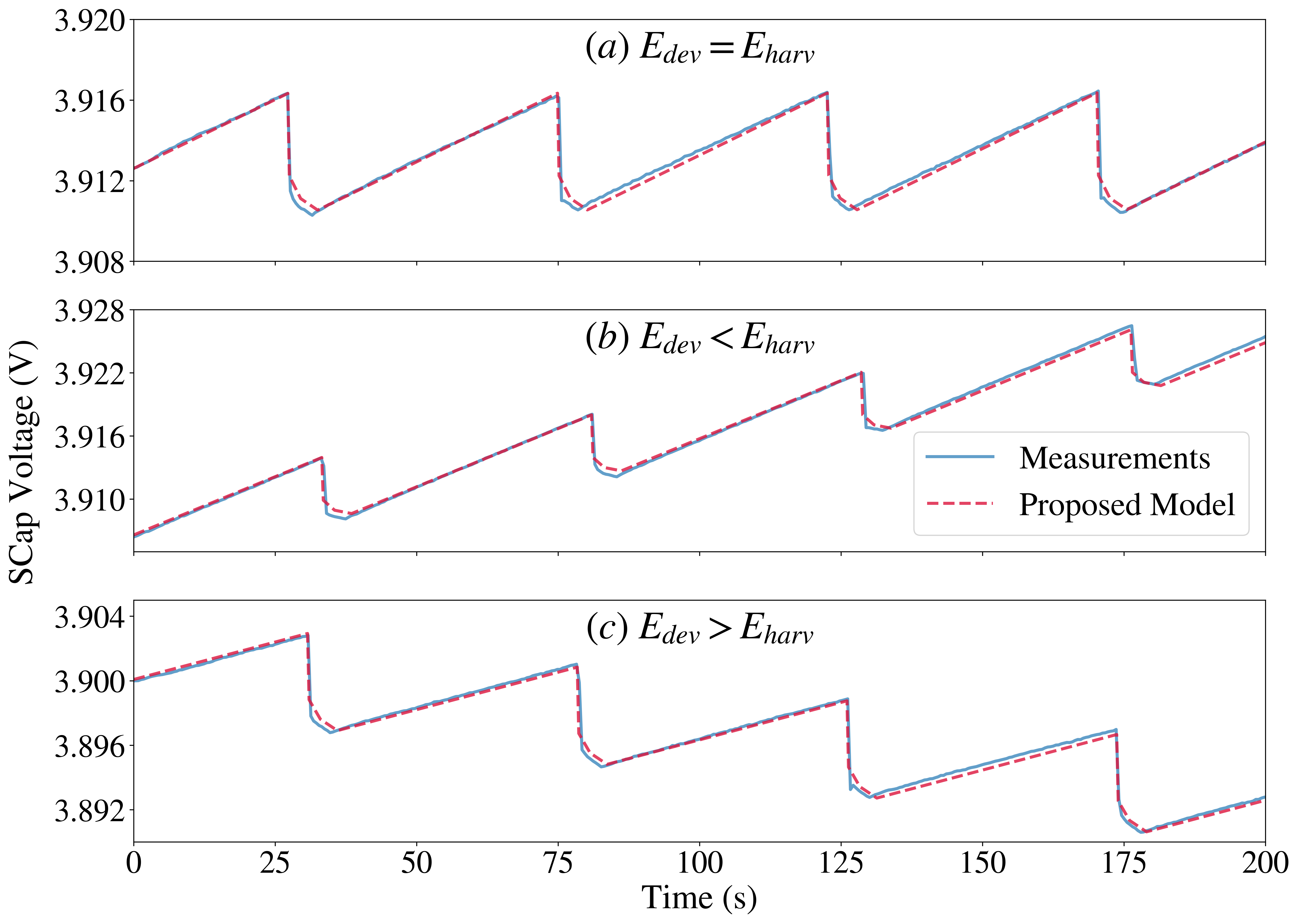}}
    \vspace{-1em}
    \caption{Performance results under the duty cycle of 500~lx.}
    \label{fig:SCapV500lx}
    \vspace{-1.5em}
\end{figure}

Results for 500~lx illumination are shown in Fig.~\ref{fig:SCapV500lx}, with a node duty cycle of 47.6~s. 
For instance,  Fig.~\ref{fig:SCapV500lx}~(b), changes in illumination (to 600~lx), representing a case in which $E_{dev} < E_{harv}$, leading to an increasing trend in the \ac{SCap} charge.
\begin{figure}[t]
    \vspace{-0.1em}
    \centerline{\includegraphics[scale=.26]{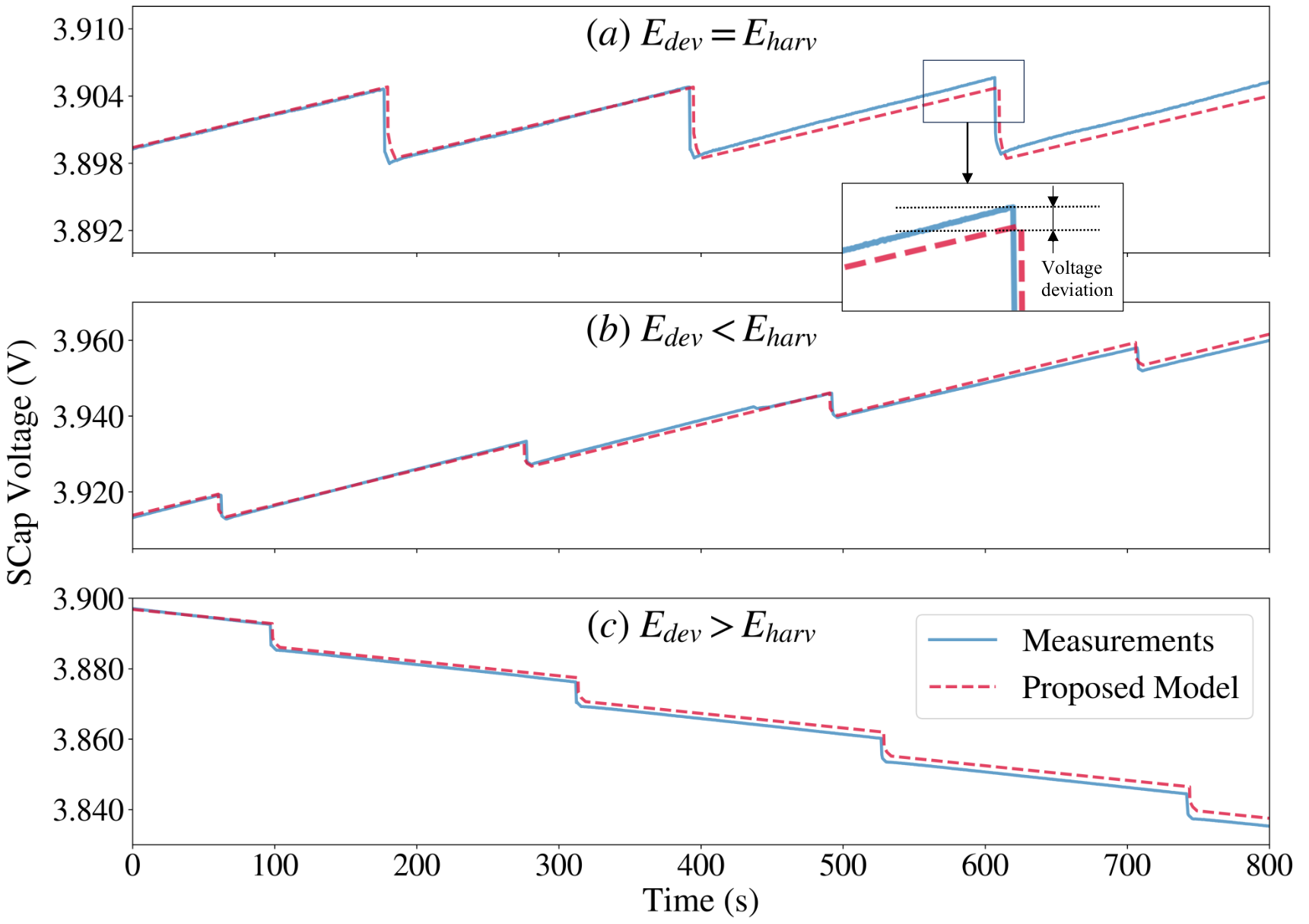}}
    \vspace{-1em}
    \caption{Performance results under the duty cycle of 300~lx.}
    \label{fig:SCapV300lx}
    \vspace{-1.6em}
\end{figure}
Figure~\ref{fig:SCapV300lx} depicts results under 300~lx, where harvested power closely matches the node's sleep energy requirements, leading to an extended sleep interval of 209.9~s (duty cycle: 215~s). Due to the need of longer energy accumulation periods, light variations significantly impact the \ac{SCap} charge, causing a slight divergence between simulation and measurements (Figs.~\ref{fig:SCapV300lx}~(a)~and~(c)). Nevertheless, the observed error remains minimal (0.0012~V over 800~s and expected 0.0454~V over 8~hours) representing around 1\% of the stored energy. This minor deviation could be further reduced by refining the characterization of hardware components, such as the \ac{EHU}. 

\section{Conclusions and future work}
\label{Sec_V}

In this work, we presented an analytical model for batteryless IoT sensor nodes powered by ambient energy harvesting. The model integrates energy harvesting, power management, and supply stages, providing a detailed representation of the \ac{SCap} voltage over a given operational time \(T\). Experimental results under three illumination levels (300~lx, 500~lx, and 700~lx) confirm that the model accurately captures \ac{SCap}'s energy behavior across sensing, communication, and sleeping states. Simulations closely match measurements from our batteryless prototype, demonstrating the model’s robustness to illumination variations, with estimated deviations around 1\% over an eight-hour operation. Future work could focus on leveraging this model for adaptive sleep mechanisms and its integration into digital twins to improve resource utilization and reliability in sustainable IoT systems.

\vspace{-0.2em}
\bibliographystyle{ieeetr}
\bibliography{bibliography}

\end{document}